\documentclass[twocolumn,amsmath,amssymb]{revtex4}
\usepackage{graphicx}
\usepackage{dcolumn}
\usepackage{color}
\usepackage{scrextend}
\usepackage{subfig}
\usepackage{bm}
\usepackage{amsmath,amssymb,graphicx}
\usepackage{epsfig}
\usepackage{enumerate}
\usepackage{array}
\usepackage{multirow}

\begin{document}
\title
{Weak-coupling limits of the quantum Langevin equation for an oscillator}
\author{Aritra Ghosh$^1$\footnote{ag34@iitbbs.ac.in} and Sushanta Dattagupta$^{2,3}$\footnote{sushantad@gmail.com} }
\affiliation{$^{1}$School of Basic Sciences, Indian Institute of Technology Bhubaneswar, Jatni, Khurda, Odisha 752050, India\\
$^{2}$National Institute of Technology Durgapur, Mahatma Gandhi Road, Durgapur, West Bengal 713209, India\\
$^{3}$Sister Nivedita University, New Town, Block DG 1/2, Action Area I,  Kolkata, West Bengal 700156, India}
\vskip-2.8cm
\date{\today}
\vskip-0.9cm

\vspace{5mm}


\begin{abstract}
The quantum Langevin equation as obtained from the independent-oscillator model describes a strong-coupling situation, devoid of the Born-Markov approximation that is employed in the context of the Gorini–Kossakowski–Sudarshan–Lindblad equation. The question we address is what happens when we implement such
`Born-Markov'-like approximations at the level of the quantum Langevin equation for a harmonic oscillator which carries a noise term satisfying a fluctuation-dissipation theorem. In this backdrop, we also comment on the rotating-wave approximation.
\end{abstract}

\maketitle

\section{Introduction}
Open quantum systems are ubiquitous \cite{Weiss}, and can be taken to be modeled as a quantum system of interest which interacts with some environment. There are mainly two distinct approaches towards tackling such systems, namely, the method of the quantum Langevin equation (QLE) \cite{QLE,QLE2} which involves dealing with effective Heisenberg equations describing the `reduced' dynamics of the system of interest, while the other approach involves the density operator, with dynamics described by a suitable master equation (see \cite{purisdgbook} and references therein). Although the latter approach has found use in the description of a wide variety of systems, including those in condensed-matter physics, atomic physics, and quantum optics, it is typically derived under the so-called Born and Markov approximations, which are in a sense, `weak-coupling' approximations. The Born approximation essentially asserts that the density matrix at all times is factorizable as
\begin{equation}\label{borncond}
\rho(t) = \rho_{\rm S}(t) \otimes \rho_B,
\end{equation} where the subscripts `S' and `B' denote the system and the bath. It is essentially a weak-coupling assumption, because it requires that the system-bath coupling does not alter the state of the bath, or the bath eigenstates. In addition, one has the Markov approximation which asserts that the bath correlations die over time rather rapidly, i.e, the bath has a `fast dynamics'. Thus, the dynamics of the density matrix depends only on its present state and not on the past. The resulting master equation takes the form of a Gorini–Kossakowski–Sudarshan–Lindblad (GKSL) equation \cite{GKSL} which has the property of being trace-preserving and completely positive for any initial condition. There are of course, other kinds of master equations, such as those of the Redfield kind \cite{Red1,Red2,Red3}, which under certain approximations, may reduce to the GKSL form. \\

The alternate route towards studying open quantum systems is provided by the QLE, which, when the system is a single particle of mass \(m\) and moving in some potential \(V(x)\) (later set to be a harmonic potential), reads
 \begin{equation}\label{qle}
m \ddot{x}(t) + \int_{0}^t \mu(t - t') \dot{x}(t') dt' + V'(x) = F(t).
\end{equation} It is an integro-differential equation for the system's position operator \(x = x(t)\). The function \(\mu(t)\) is called the dissipation kernel, which describes the memory effects involved in the drag force (proportional to the velocity \(\dot{x}\)) that the system experiences and \(F(t)\) is an operator-valued random force, called the noise and is described by its statistical properties \cite{QLE,lho1}. Both \(\mu(t)\) and \(F(t)\) are usually determined from a microscopic model which is the usual starting point of the problem \cite{lho1,FordQT0}. Different physical quantities of interest can now be obtained by integrating the QLE and computing relevant correlation functions, e.g., the kinetic energy is obtained from the equal-time correlation \(\langle \dot{x}(t) \dot{x}(t)\rangle\), where the averaging is performed over all possible noise realizations. This ascribes a thermal structure to such averages, since the noise originates from the heat bath at thermal equilibrium. For definiteness, in the subsequent analysis, we take the heat bath to be composed of an infinite number of independent oscillators and the system's coupling to the heat bath as bilinear; this is the famous independent-oscillator model \cite{QLE,FordQT0}, also called the Feynman-Vernon model \cite{FV} or the Caldeira-Leggett model \cite{CL}. \\

In this short note, we shall argue that the QLE as obtained from the independent-oscillator model does correspond to a strong-coupling scenario, unlike the case of Born-Markov master equations which are obtained in a weak-coupling setting. We then discuss the implementation of such approximations at the level of the QLE when the system is a harmonic oscillator, which then gives equal-time correlation functions (in the steady state) identical to those obtained from the master equation \cite{GSA}. In particular, we describe the weak-coupling limits of the steady-state position and velocity autocorrelation functions for the dissipative oscillator, as obtained from the QLE. In this context, we comment on the rotating-wave approximation (RWA).
 
\section{Quantum Langevin equation}\label{PreSec}
In this section, we revisit the subtleties associated with the derivation of the QLE, starting from the independent-oscillator model. The total Hamiltonian reads as \(H = H_{\rm S} + H_{\rm B} + H_{\rm SB}\), where 
\begin{eqnarray}
&&H_\mathrm{S}=\frac{p^2}{2m}+ V(x),\label{HS}\\
&&H_\mathrm{B}=\sum_{j=1}^{N} \Bigg[\frac{p_j^2}{2m_j}+\frac{1}{2}m_j\omega_j^2q_j^2\Bigg],\label{HB}\\
&&H_\mathrm{SB}=-x\sum_{j=1}^Nc_j q_j+x^2\sum_{j=1}^N\frac{c_j^2}{2m_j\omega_j^2}, \label{HSB}
\end{eqnarray}
are respectively, the Hamiltonians describing the system, the heat bath, and their interactions. Here $c_j$ denotes the coefficient of linear coupling between the coordinate of the particle and that of the $j$-th oscillator of the heat bath. The symbols \(x\) and \(p\) denote the position and momentum operators of the system, while \(\{q_j\}\) and \(\{p_j\}\) are the corresponding operators for the heat-bath oscillators. Naturally, we have the commutator algebra: 
\begin{equation}
[x,p] = i\hbar, \hspace{5mm} [q_j,p_k] = i \hbar \delta_{j,k},
\end{equation} with all other commutators vanishing. Upon solving the Heisenberg equations for the bath variables, and then substituting them into the equation of motion of the system, one obtains \cite{bez,physicaA,hist}
\begin{equation}\label{qle0}
m \ddot{x}(t) + \int_{0}^t \mu(t - t') \dot{x}(t') dt' + \mu(t) x(0) + V'(x) = g(t),
\end{equation}
where we have defined
\begin{eqnarray}
\mu(t) &=& \Theta(t) \sum_{j = 1}^N \frac{c_j^2}{m_j \omega_j^2} \cos (\omega_j t), \label{1}\\
g(t)&=&\sum_{j=1}^{N} c_j \Bigg[q_j(0)\cos(\omega_jt)+\frac{p_j(0)}{m_j\omega_j}\sin(\omega_jt)\Bigg]. \label{noise}
\end{eqnarray}
Both \(\mu(t)\) and \(g(t)\) depend upon the microscopic parameters of the heat bath, while \(g(t)\) also depends upon the initial conditions of the heat-bath oscillators. Clearly, \(\mu(t)\) plays the role of a `friction kernel', i.e., it determines how the drag force experienced by the system depends upon velocities of the past; a friction kernel which is not proportional to a delta function shall therefore indicate towards a non-Markovian drag force. Note that the step-function appearing in Eq. (\ref{1}) enforces causality.  \\

To understand the physical meaning of \(g(t)\), we need to have a closer look at the initial preparation of the system and the heat bath. While we can choose the initial conditions of the system from any suitable distribution function (so that the uncertainty principle is met), we take the heat bath to be in a state of thermal equilibrium for all times, including the initial instant \(t = 0\). Thus, the initial density matrix of the heat bath reads
\begin{equation}\label{rho011}
\rho_{\rm B}(0) = \frac{e^{\bigg[{-\beta \sum_{j=1}^N \bigg[ \frac{p_j^2(0)}{2m_j} + \frac{m_j \omega_j^2 q_j^2(0)}{2} \bigg]}\bigg]}}{\Lambda_0},
\end{equation} where \(\Lambda_0\) is a suitable normalizing factor. The full initial state can be taken to be \(\rho(0) = \rho_{\rm S}(0) \otimes \rho_{\rm B}(0)\), i.e., the system and the heat bath are decoupled at the initial instant. Then, upon averaging with respect to \(\rho_{\rm B}(0)\), it follows that \(g(t)\) is a Gaussian noise, and therefore Eq. (\ref{qle0}) now appears like a QLE, except for the bizarre term \(\mu(t) x(0)\), often referred to as the `initial-slip term' \cite{bez,hist}. Thus, the independent-oscillator model with the above-mentioned initial preparation does not exactly lead to the QLE [Eq. (\ref{qle})], although it comes close. As it turns out, the exact form of Eq. (\ref{qle}) emerges for a quite different initial preparation, which corresponds to the following initial density matrix for the bath (+ interactions): 
\begin{equation}\label{rhobsb}
\rho_{\rm B + SB}(0) = \frac{e^{\Bigg[{-\beta \sum_{j=1}^N \bigg[ \frac{p_j^2(0)}{2m_j} + \frac{m_j \omega_j^2}{2} \Big(q_j(0) -\frac{c_jx(0)}{m_j\omega_j^2}\Big)^2  \bigg]}\Bigg]}}{\overline{\Lambda}_0},
\end{equation}
for a different normalizing factor, \(\overline{\Lambda}_0\). Averaging with respect to the above-mentioned distribution function \cite{bez,physicaA,hist,SDG2}, it is easy to see that \(f(t) \equiv g(t) - \mu(t) x(0)\), i.e., 
\begin{equation}\label{ft}
f(t) = \sum_{j=1}^{N} c_j \Bigg[\bigg(q_j(0) - \frac{c_j x(0)}{m_j \omega_j^2}\bigg)\cos(\omega_jt)+\frac{p_j(0)}{m_j\omega_j}\sin(\omega_jt)\Bigg]
\end{equation}
 is a Gaussian noise, satisfying the same statistical properties as \(g(t)\) [the latter is averaged over Eq. (\ref{rho011})]. One now has the QLE given in Eq. (\ref{qle}), with the identification \(F(t) = f(t) \equiv g(t) - \mu(t) x(0)\). Thus, in order to obtain the QLE [Eq. (\ref{qle})] from the independent-oscillator Hamiltonian, one has the resort to an initial preparation of the system and the bath such that the initial density matrix of the system and the heat bath taken together cannot be totally factorized into a system density matrix times the density matrix of the `isolated' heat bath with no inter-correlations. In fact, as one can easily verify, at \(t = 0\), there are non-trivial correlations between the system and the bath, if averaged over \(\rho_{\rm B + SB}(0)\). Since the Born approximation invariably implies Eq. (\ref{borncond}) for not just \(t = 0\), but also for subsequent times, one may conclude that the QLE does not correspond to a weak-coupling situation, as taken by the Born approximation. It should be remarked that classical noises with their moments obtained by averaging over a conditional distribution such as Eq. (\ref{rhobsb}) were obtained by Zwanzig \cite{Zwanzig}. Finally, it may be emphasized that the full density matrix at the initial instant and also for subsequent times is given by the Gibbs canonical state
 \begin{equation}
 \rho = \frac{e^{-\beta H}}{Z}, 
 \end{equation} where \(H = H_{\rm S} + H_{\rm B} + H_{\rm SB}\) and is clearly not factorizable as a direct product of density matrices for the system and the bath.

\subsection{Density of states of the heat bath}
In the (most reasonable) limit \(N \rightarrow \infty\), one may replace the sums over the index \(j\) with an integral as
\begin{equation}\label{sum-integral}
\sum_{j=1}^N \rightarrow N \int_0^\infty d\omega g(\omega),
\end{equation} where \(g(\omega)\) is the density of states of the heat bath. To this end, let us consider a rather simple situation where all the heat-bath oscillators have the same mass \(m_j = \widetilde{m}\), and moreover the coupling constants \(\{c_j\}\) are all equal and scale as inverse square-root of the number of oscillators, i.e., we have \(c_j = \widetilde{c}/\sqrt{N}\). In this case, upon replacing the sum with an integral, Eq.~(\ref{1}) reads
\begin{equation}\label{mutog}
\mu(t) = \Theta(t) \frac{\widetilde{c}^2}{\widetilde{m}} \int_0^\infty d\omega~\frac{g(\omega)}{\omega^2} \cos (\omega t).
\end{equation} 
We shall call the heat bath an Ohmic bath if \(g(\omega) \sim \omega^2\), with the frequencies restricted to a certain high-frequency cutoff, i.e., one has \cite{purisdgbook}
\begin{eqnarray}
g(\omega) &=& \frac{3 \omega^2}{\Omega^3};\hspace{6mm} \omega < \Omega, \label{gomegaohmic} \\
&=& 0; \hspace{11mm} \omega > \Omega.  \nonumber
\end{eqnarray} 
Substituting the above choice in Eq.~(\ref{mutog}), we find
\begin{equation}\label{mutgenohmic}
\mu(t) = \Theta(t)\frac{3 \widetilde{c}^2}{\widetilde{m} \Omega^3} \frac{ \sin (\Omega t)}{t}.
\end{equation}
We now consider the limit of very large \(\Omega\). Using $\delta(x)=\lim_{\Omega \to \infty}\sin(\Omega x)/(\pi x)$ in Eq. (\ref{mutgenohmic}), it follows that
\begin{equation}\label{mutohmic}
\mu(t) \approx 2 m \gamma \delta(t),
\end{equation}
where one has
\begin{equation}\label{gammamicro}
\gamma \equiv \frac{3 \pi \widetilde{c}^2}{2 m \widetilde{m} \Omega^3},
\end{equation} which would also go to zero, unless we take $\widetilde{c}$ to scale as $\Omega^{3/2}$. Eq. (\ref{mutohmic}) is referred to as the `strict' Ohmic case. It corresponds to a situation where the drag force on the system is instantaneous, i.e., it depends upon the velocity of the particle at time \(t\) and not on the history of the velocity. 

\subsection{Spectral properties of the noise \(f(t)\)}
Since the noise is defined in Eq. (\ref{ft}), and the initial density matrix describing the heat bath and the interactions between the system and the bath is given by Eq. (\ref{rhobsb}), one finds that \(\langle f(t) \rangle = 0\), where the angled brackets denote an averaging over the Gaussian distribution given in Eq. (\ref{rhobsb}). This means averaging over all possible noise realizations. Further, from Eq. (\ref{rhobsb}), one has the second moments: 
\begin{equation}\label{qq}
\bigg\langle \bigg(q_j(0) - \frac{c_j x(0)}{m_j \omega_j^2}\bigg)^2 \bigg\rangle = \frac{\hbar}{2 m_j \omega_j} \coth \bigg(\frac{\hbar \omega_j}{2 k_B T}\bigg), 
\end{equation}

\begin{equation}\label{pp}
\langle p_j^2(0) \rangle = \frac{\hbar m_j \omega_j}{2} \coth \bigg(\frac{\hbar \omega_j}{2 k_B T}\bigg),
\end{equation} 

\begin{equation}\label{qp}
\bigg\langle \bigg(q_j(0) - \frac{c_j x(0)}{m_j \omega_j^2}\bigg) p_j(0) \bigg\rangle = - \bigg\langle p_j(0) \bigg(q_j(0) - \frac{c_j x(0)}{m_j \omega_j^2}\bigg) \bigg\rangle  = \frac{i \hbar}{2}, \end{equation} with the others vanishing. Notice that in the classical limit, i.e., for \(\hbar \rightarrow 0\), these correlations conform to the results expected from the classical equipartition theorem, because \(\coth x \approx 1/x\) for \(x \rightarrow 0\). Now, using Eqs. (\ref{qq})-(\ref{qp}), one has 
\begin{equation}
\langle  f(t) f(t') \rangle =  \sum_{j=1}^N \frac{c_j^2}{2 m_j\omega_j} \hbar \coth \bigg(\frac{\hbar \omega_j}{2k_B T}\bigg)  e^{i \omega_j (t-t')}. 
\end{equation}
Introducing the spectral density function as \cite{Weiss}
\begin{equation}\label{Jdef}
J(\omega) \equiv \frac{\pi}{2} \sum_{j=1}^N \frac{c_j^2}{m_j \omega_j} \delta (\omega - \omega_j),
\end{equation}
we have in the continuum limit, the following expression:
\begin{equation}
\langle  f(t) f(t') \rangle = \frac{\hbar}{\pi}  \int_0^\infty d\omega J(\omega) \coth \bigg(\frac{\hbar \omega}{2k_B T}\bigg)  e^{i \omega (t-t')}. 
\end{equation}
The spectral properties of the noise are therefore given by
\begin{eqnarray}
\langle \lbrace f(t), f(t') \rbrace \rangle &=& \frac{2\hbar}{\pi}\int_{0}^{\infty}d\omega J(\omega) \coth\bigg(\frac{\hbar\omega}{2k_BT}\bigg) \nonumber \\
&&~~~~~~~~ \times \cos \lbrack \omega(t-t')\rbrack,  \label{symmetricnoisecorrelation1} \\
\langle \lbrack f(t), f(t') \rbrack \rangle &=& \frac{2\hbar}{i\pi}\int_{0}^{\infty}d\omega J(\omega) \sin\lbrack \omega(t-t')\rbrack . 
\label{noisecommutator1}
\end{eqnarray} 
The reader may note that the (spectral) density function \(J(\omega)\) is closely related to the density of states \(g(\omega)\) of the heat bath. From Eqs. (\ref{1}) and (\ref{Jdef}), we have
\begin{equation}
\mu(t) = \Theta(t) \frac{2}{\pi} \int_0^\infty \frac{J(\omega)}{\omega} \cos (\omega t) d\omega.
\end{equation} Comparing this with Eq. (\ref{mutog}), we have
\begin{equation}
J(\omega) \sim \frac{\pi}{2} \frac{\widetilde{c}^2}{\widetilde{m}} \frac{g(\omega)}{\omega} ,
\end{equation} and therefore for (strict) Ohmic dissipation, \(J(\omega) = m \gamma \omega\) for \(\omega \in [0,\infty)\). 

\subsection{Callen-Welton fluctuation-dissipation theorem}
For explicit calculation of correlation functions, we resort to the simple case where the system is a quantum harmonic oscillator, i.e., we make the choice \(V(x) = m \omega_0^2 x^2/2\) in Eq. (\ref{HS}). Then Eq. (\ref{qle}) is linear and can be solved exactly using a Green's function as \(\widetilde{x}(\omega) = \alpha(\omega) \widetilde{f}(\omega)\), where \(\alpha (\omega)\) is called the generalized susceptibility, given by (here `tilde' denotes Fourier transform)
\begin{equation}\label{alphadef11}
\alpha(\omega) = \frac{1}{m(\omega_0^2 - \omega^2) - i \omega \widetilde{\mu}(\omega)}.
\end{equation} Notice that \(\alpha(\omega)\) is just the Fourier transform of the Green's function in the time domain. It is noteworthy that there are subtle issues involving the quantum Langevin equation if one goes beyond harmonic oscillators \cite{nonlin1,nonlin2}; we do not consider such cases here. In the present case where the system is a harmonic oscillator, the steady-state correlation functions can be computed using the fluctuation-dissipation theorem of the Callen-Welton kind, which we now state without proof (see \cite{Weiss,callenwelton,Kubo,case,bb} for details):
\begin{widetext}
\begin{eqnarray}
C_x(t-t')=\frac{1}{2} \langle x(t) x(t') + x(t') x(t) \rangle = \frac{\hbar}{\pi} \int_0^\infty {\rm Im} [\alpha(\omega + i 0^+)]  \coth \bigg(\frac{\hbar \omega}{2k_B T}\bigg) \cos [\omega(t-t')] d\omega.  \label{CW}
\end{eqnarray} One may find the velocity-autocorrelation function by differentiating Eq. (\ref{CW}) with respect to \(t\) and \(t'\), giving
\begin{eqnarray}
C_{\dot{x}}(t-t') = \frac{1}{2} \langle \dot{x}(t) \dot{x}(t') + \dot{x}(t') \dot{x}(t) \rangle = \frac{\hbar}{\pi} \int_0^\infty \omega^2 {\rm Im} [\alpha(\omega + i 0^+)]  \coth \bigg(\frac{\hbar \omega}{2k_B T}\bigg) \cos [\omega(t-t')] d\omega.  \label{CW1}
\end{eqnarray}
\end{widetext} 
In the context of the quantum counterpart of energy equipartition theorem \cite{jarzy2,jarzy3,agmb} (see \cite{kaur} for a three-dimensional generalization), one can easily show that the following are probability distributions:
\begin{equation}\label{pk}
P_k (\omega) = \frac{2 m \omega }{ \pi} {\rm Im} [\alpha(\omega + i 0^+)],
\end{equation} and 
\begin{equation}\label{pp}
P_p (\omega) = \frac{2 m \omega_0^2 }{ \omega \pi } {\rm Im} [\alpha(\omega + i 0^+)],
\end{equation} i.e., \(P_{k,p} (\omega) \geq 0\), \(\forall \omega \in [0,\infty)\) and they are also normalized in this interval (see \cite{jarzy2,kaur,agmb} for proof). Thus, the steady-state correlation functions read
\begin{equation}\label{Cxweak}
C_x(\tau) = \frac{\hbar }{2m \omega_0^2} \int_0^\infty \omega P_p(\omega) \coth \bigg(\frac{\hbar \omega}{2k_B T}\bigg) \cos (\omega\tau) d\omega,
\end{equation}
\begin{equation}\label{Cxdotweak}
C_{\dot{x}}(\tau) = \frac{\hbar }{2m} \int_0^\infty \omega P_k(\omega) \coth \bigg(\frac{\hbar \omega}{2k_B T}\bigg) \cos (\omega\tau) d\omega,
\end{equation}
where \(\tau = t - t'\). 

\section{Born \& Markov approximations}

\subsection{Born approximation}
We now discuss the Born approximation which is essentially a weak-coupling limit in the sense that the density matrix of the system-plus-bath is factorizable for all times. This is not true for the QLE, for we have seen that for consistency, the system and the heat bath must be coupled even at the initial instant and the density matrix is not factorizable. A weak-coupling limit would then mean taking the system-bath coupling to be sufficiently small so that density matrix may be factorized, as in the Born approximation. \\

\begin{figure}
\subfloat[\(\Gamma = 1 \)\label{fig:test1}]
  {\includegraphics[width=0.8\linewidth]{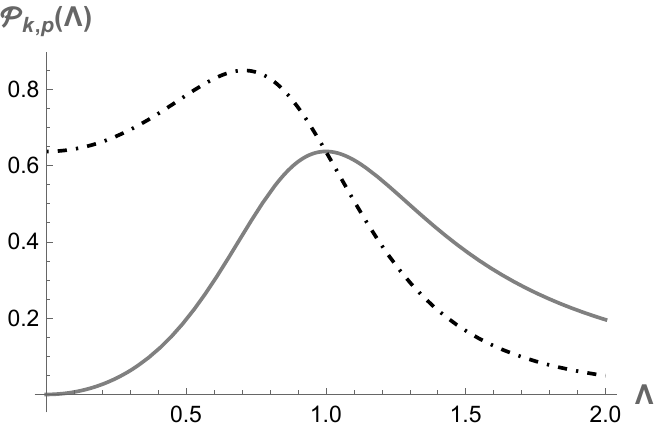}}\hfill
\subfloat[\(\Gamma = 0.5\).\label{fig:test2}]
  {\includegraphics[width=0.8\linewidth]{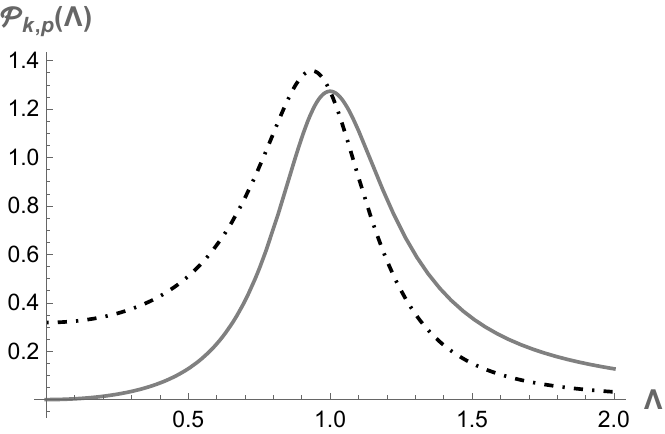}}\hfill
\subfloat[\(\Gamma = 0.125 \).\label{fig:test3}]
  {\includegraphics[width=0.8\linewidth]{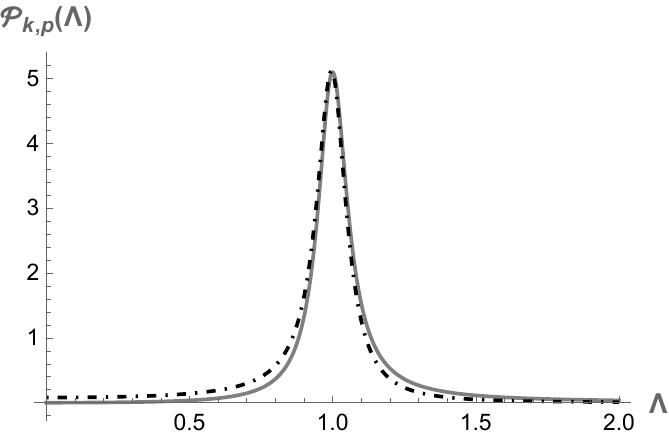}} \hfill
  \subfloat[\(\Gamma = 0.0125 \).\label{fig:test4}]
  {\includegraphics[width=0.8\linewidth]{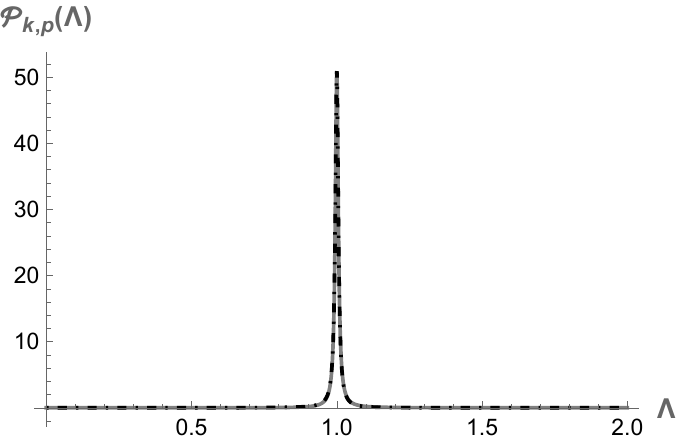}}
\caption{The (dimensionless) distribution functions \(\mathcal{P}_k(\Lambda)\) (gray-solid) and \(\mathcal{P}_p(\Lambda)\) (dotted-dashed) for different values of \(\Gamma\).}
\end{figure}

To this end, let us revisit the simple case of strict Ohmic dissipation for which \(\mu(t) = 2m \gamma \delta(t)\). The corresponding susceptibility is \(\alpha(\omega) = [m(\omega_0^2 - \omega^2) - i m\omega \gamma]^{-1}\), which means
\begin{eqnarray}
 {\rm Im} [\alpha(\omega)] &=& \frac{1}{m} \frac{\omega \gamma}{(\omega_0^2 - \omega^2)^2 + (\omega \gamma)^2} \nonumber \\
 &=& \frac{1}{m \omega_0^2} \bigg(\frac{\Lambda \Gamma}{(1 - \Lambda^2)^2 + (\Lambda \Gamma)^2}\bigg),
\end{eqnarray} where \(\Gamma = \gamma/\omega_0\) and \(\Lambda = \omega/\omega_0\), with the function within the large parenthesis after the second equality being dimensionless. Thus, we can define the dimensionless functions \(\mathcal{P}_k(\Lambda) = \omega_0 P_k(\omega/\omega_0)\) and \(\mathcal{P}_p(\Lambda) = \omega_0 P_p(\omega/\omega_0)\), which read
\begin{equation}
\mathcal{P}_k(\Lambda) = \frac{2}{\pi} \frac{\Lambda^2 \Gamma}{(1 - \Lambda^2)^2 + (\Lambda \Gamma)^2}= \frac{2}{\pi} \frac{ \Gamma}{(\Lambda - \Lambda^{-1})^2 +  \Gamma^2},
\end{equation}
\begin{equation}
\mathcal{P}_p(\Lambda) = \frac{2}{\pi} \frac{ \Gamma}{(1 - \Lambda^2)^2 + (\Lambda \Gamma)^2} =  \frac{2}{\pi \Lambda^2} \frac{ \Gamma}{(\Lambda - \Lambda^{-1})^2 +  \Gamma^2}.
\end{equation}
We may now smoothly consider the limit \(\Lambda \rightarrow 0\), i.e., \(\gamma << \omega_0\), for which the Lorentzian factors above reduce to delta functions as
\begin{equation}\label{limitPequation}
\lim_{\Gamma \rightarrow 0} \mathcal{P}_{k,p}(\Lambda) \approx \delta(\Lambda - 1), \quad \Lambda \geq 0. 
\end{equation} The delta function peaks at \(\Lambda = 1\) which is \(\omega = \omega_0\). Thus, it is easy to show from Eqs. (\ref{Cxweak}) and (\ref{Cxdotweak}) that
\begin{equation}\label{Cxweak1}
C_x(\tau)\big|_{\gamma \rightarrow 0^+} = \frac{\hbar }{2m \omega_0}   \coth \bigg(\frac{\hbar \omega_0}{2k_B T}\bigg) \cos (\omega_0\tau),
\end{equation}
\begin{equation}\label{Cxdotweak1}
C_{\dot{x}}(\tau)\big|_{\gamma \rightarrow 0^+} = \frac{\hbar \omega_0 }{2m} \coth \bigg(\frac{\hbar \omega_0}{2k_B T}\bigg) \cos (\omega_0\tau).
\end{equation}
The correlation functions are oscillatory in \(\tau\), with frequency \(\omega_0\) which is the characteristic frequency of the system (the harmonic oscillator). This is rather expected because in the weak-coupling limit, the dynamics of the open system is dominated by the characteristic timescale \(\omega_0^{-1}\) and that there is no appreciable broadening of the frequencies due to dissipation. For the purpose of illustration, we have plotted \(\mathcal{P}_k(\Lambda)\) and \(\mathcal{P}_p(\Lambda)\) for different values of \(\Gamma\) in Figs. (\ref{fig:test1}), (\ref{fig:test2}), (\ref{fig:test3}), and (\ref{fig:test4}); one clearly observes that both the distribution functions approach delta-function behavior as \(\Gamma\) takes smaller values, as expected from Eq. (\ref{limitPequation}).\\

Putting \(\tau = 0\), one obtains the kinetic and potential energies of the dissipative oscillator in the weak-coupling regime as
\begin{equation}\label{Ekfinal}
E_k\big|_{\gamma \rightarrow 0^+} \equiv m C_{\dot{x}}(0)\big|_{\gamma \rightarrow 0^+} = \frac{\hbar \omega_0 }{2} \coth \bigg(\frac{\hbar \omega_0}{2k_B T}\bigg), 
\end{equation}
\begin{equation}\label{Epfinal}
E_p\big|_{\gamma \rightarrow 0^+} \equiv m \omega_0^2 C_{x}(0)\big|_{\gamma \rightarrow 0^+} = \frac{\hbar \omega_0 }{2} \coth \bigg(\frac{\hbar \omega_0}{2k_B T}\bigg), 
\end{equation} which are just the well-known weak-coupling results for an oscillator in contact with a heat bath \cite{agmb}. Note that for arbitrary coupling strengths, the thermally-averaged kinetic and potential energies of the oscillator are not equal \cite{virial}. This however, does not imply that the general structure of the virial theorem does not hold just because the thermally-averaged kinetic and potential energies are unequal; it happens to acquire bath-induced quantum-correction terms which render the thermally-averaged kinetic and potential energies unequal for strong system-bath coupling, where a non-Markovian (quantum) noise is present. In the weak-coupling limit, the equality between the thermally-averaged kinetic and potential energies is restored, as also seen from Figs. (\ref{fig:test1}), (\ref{fig:test2}), (\ref{fig:test3}), and (\ref{fig:test4}).

\subsection{Markov approximation}
The Markov approximation involves the assumption that the correlations associated with the bath fall off rapidly. At the level of the independent-oscillator model and the QLE that follows from it, this assumption translates to the fact that the bath cutoff frequency \(\Omega\) which appears in Eq. (\ref{gomegaohmic}) is quite large. This is not surprising because the timescale \(\Omega^{-1}\) characterizes the time over which the density of states falls off. Quite obviously, the friction kernel [Eq. (\ref{mutgenohmic})] falls off too, over the same timescale and reduces to a delta-function, characterizing Markovian, i.e., memoryless damping. \\

 Now, taking \(\gamma\) as defined in Eq. (\ref{gammamicro}) to be finite, let us have a look at Eq. (\ref{symmetricnoisecorrelation1}). Since the bath now has a fast dynamics, in the sense that the correlations must fall off rapidly over the timescale \(\Omega^{-1}\) which tends to zero, we may approximate the integral in Eq. (\ref{symmetricnoisecorrelation1}) around the point \(\omega = \omega_0\) as

 \begin{equation}\label{fMarkov}
 \langle \lbrace f(t), f(t') \rbrace \rangle \approx 2m\gamma\hbar \omega_0  \coth\bigg(\frac{\hbar\omega_0}{2k_BT}\bigg) \delta(t-t'),
 \end{equation} i.e., the noise now appears to be delta-correlated, consistent with the Markov assumption and we have put \(J(\omega) = m \gamma \omega\), consistent with strict Ohmic dissipation (because \(\mu(t) \sim \delta(t)\)). We may express this as \( \langle \lbrace f(t), f(t') \rbrace \rangle \approx  \Gamma \delta(t-t')\), with \(\Gamma = 2m\gamma\hbar \omega_0 \coth\big(\frac{\hbar\omega_0}{2k_BT}\big)\). Therefore, the QLE for the oscillator reads
 \begin{equation}\label{qlemarkov}
  \ddot{x}(t) + \gamma \dot{x}(t) + \omega_0^2 x(t) = \frac{f(t)}{m},
 \end{equation} where \(f(t)\) is a Markovian noise, satisfying Eq. (\ref{fMarkov}). In the classical limit, i.e., for \(\hbar \rightarrow 0\), we get \(\Gamma \rightarrow 4 m \gamma k_B T\) which is the expected expression. It should be remarked here that Eq. (\ref{qlemarkov}) is equivalent to the weak-coupling Langevin equations obtained in \cite{GSA}. \\
 
 Let us finally examine the position and velocity autocorrelation functions in this limit. Integrating Eq. (\ref{qlemarkov}) directly gives \cite{chandra,sdg1}
 \begin{equation}\label{xtsol}
 x(t) = \frac{1}{\omega_+ - \omega_-} \int_0^t dt' \big[e^{\omega_+(t - t')} - e^{\omega_-(t-t')}\big] \frac{f(t')}{m},
 \end{equation} where
 \begin{equation}
 \omega_\pm = -\frac{\gamma}{2} \pm \frac{1}{2} \sqrt{\gamma^2 - 4 \omega_0^2},
 \end{equation} and in Eq. (\ref{xtsol}), we have ignored the terms dependent on the initial conditions \(x(0)\) and \(\dot{x}(0)\), for they are inconsequential when it comes to determining the long-time behavior. 
 \begin{widetext}
 Thus, we find the equal-time autocorrelation function for the position in the long-time limit as
 \begin{equation}
\lim_{t \rightarrow \infty} \langle x(t)^2 \rangle = \frac{\Gamma}{m^2(\omega_+ - \omega_-)^2} \lim_{t \rightarrow \infty} \int_0^t \int_0^t dt' dt'' \big[e^{\omega_+(t - t')} - e^{\omega_-(t-t')}\big] \big[e^{\omega_+(t - t'')} - e^{\omega_-(t-t'')}\big] \delta(t'-t''),
 \end{equation}
 where we have also used Eq. (\ref{fMarkov}). 
 \end{widetext}
 Performing these integrals and substituting for \(\Gamma\), one easily finds
\begin{equation}\label{xxeq1}
\lim_{t \rightarrow \infty} \langle x(t)^2 \rangle = \frac{\hbar}{m\omega_0} \coth \bigg(\frac{\hbar \omega_0}{2 k_B T}\bigg),
\end{equation} meaning that the averaged potential energy agrees with Eq. (\ref{Epfinal}). 
\begin{widetext}Similarly, one has for large times, the equal-time autocorrelation for the velocity as
 \begin{equation}
\lim_{t \rightarrow \infty} \langle \dot{x}(t)^2 \rangle = \frac{\Gamma}{m^2(\omega_+ - \omega_-)^2} \lim_{t \rightarrow \infty} \int_0^t \int_0^t dt' dt'' \big[\omega_+e^{\omega_+(t - t')} - \omega_-e^{\omega_-(t-t')}\big] \big[\omega_+e^{\omega_+(t - t'')} - \omega_-e^{\omega_-(t-t'')}\big] \delta(t'-t''),
 \end{equation} 
 \end{widetext}which finally gives
 \begin{equation}\label{xxeq}
\lim_{t \rightarrow \infty} \langle \dot{x}(t)^2 \rangle = \frac{\hbar\omega_0}{m} \coth \bigg(\frac{\hbar \omega_0}{2 k_B T}\bigg),
\end{equation} giving the same result as Eq. (\ref{Ekfinal}); the Markov approximation may be fully justified only when the system-bath coupling is sufficiently weak, in which case it is also justified to invoke the Born (factorization) approximation. As regards with the virial theorem mentioned briefly below Eq. (\ref{Epfinal}), it should be mentioned that a Markovian, i.e., a white noise leads to the equality between the thermally-averaged kinetic and potential energies in our model (see also, the related papers \cite{virial1,virial2}).

\subsection{Partition function in weak-coupling limit}
The (reduced) density matrix of the system is obtained by tracing over the environmental effects. The corresponding normalizing factor, namely, the reduced partition function can be evaluated using Euclidean path integrals (see \cite{Ingold} for technical details). Quite generically, for the reduced partition function one has the following expression \cite{Weiss,hangpath,SDGPath,SDG2}:
\begin{equation}
\mathcal{Z} = \frac{{\rm Tr}_{\rm S + B}\Big[e^{-\beta \big(H_{\rm S} + H_{\rm B} + H_{\rm SB}\big)}\Big]}{{\rm Tr}_{\rm B}\Big[e^{-\beta  H_{\rm B}} \Big]}.
\end{equation}
In the weak-coupling limit, one can approximate \({\rm Tr}_{\rm S + B}\big[e^{-\beta (H_{\rm S} + H_{\rm B} + H_{\rm SB})}\big] \approx {\rm Tr}_{\rm S + B}\big[e^{-\beta (H_{\rm S} + H_{\rm B})}\big] = {\rm Tr}_{\rm S}\big[e^{-\beta H_{\rm S}}\big] {\rm Tr}_{\rm B}\big[e^{-\beta H_{\rm B}}\big]\), where in the last equality, we used the fact that the system and the bath variables are independent and the corresponding Hamiltonians mutually commute. Thus, one has to a first approximation, the result that goes as
\begin{equation}
\mathcal{Z}_{\rm w.c.} \approx {\rm Tr}_{\rm S}\big[e^{-\beta H_{\rm S}}\big],
\end{equation} where w.c. denotes that the result is approximately true only in the weak-coupling limit. Furthermore, in a weak-coupling limit, the energy spectrum of a harmonic oscillator is the usual \(\epsilon_n = \hbar \omega_0 (n + 1/2)\) (bath-induced effects are negligible), meaning that the reduced partition function is \(\mathcal{Z}_{\rm w.c.} =  [2 \sinh (\beta \hbar \omega_0/2)]^{-1}\). This gives the mean energy of the oscillator to be 
\begin{equation}
E_{\rm w.c.} := - \frac{\partial}{\partial \beta} \ln \mathcal{Z}_{\rm w.c.} = \frac{\hbar \omega_0}{2} \coth \bigg(\frac{\hbar \omega_0}{2 k_B T}\bigg),
\end{equation} which is the familiar textbook result, consistent with Eqs. (\ref{Ekfinal}) and (\ref{Epfinal}). We should emphasize again that the above-mentioned result holds for the weak-coupling limit, and not in general (see \cite{agmb} and references therein for discussions on the energetics for arbitrary system-bath coupling strength).

\subsubsection{Free-particle limit}
Let us now comment on the free-particle limit of the quantum Brownian oscillator, i.e., in which we take \(\omega_0 \rightarrow 0\). In the weak-coupling approximation, the noise correlation [Eq. (\ref{fMarkov})] then reduces to \begin{equation}
\frac{\langle \{ f(t), f(t') \} \rangle}{2} =  2 m \gamma k_B T \delta(t-t'). 
\end{equation}
This appears to match exactly with the noise correlation for a classical Brownian particle, except that \(f(t)\) above is still operator-valued and we are in the quantum regime \((\hbar \neq 0)\). This is in contrast to the arbitrary-coupling case where even for a free quantum Brownian particle, the noise correlation is quite non-trivial. One may understand the reason behind the appearance of a classical-like result in the weak-coupling limit for the free particle by noticing that within the weak-coupling limit, the (reduced) canonical partition function of the system is just
\begin{equation}
\mathcal{Z}_{\rm w.c.} = \sum_k e^{-\beta \epsilon_k}, \hspace{5mm} \epsilon_k = \frac{\hbar^2 k^2}{2m}. 
\end{equation}
This agrees exactly with the classical result, and gives mean kinetic energy to be \(k_B T/2\), independent of \(\hbar\). On the other hand, for arbitrary coupling strengths, the partition function is not given by the above-mentioned expression and therefore, the equilibrium statistical mechanics of the free quantum Brownian particle for arbitrary system-bath coupling strength is quite different from that of its classical counterpart. It is very interesting to note that classically, the presence of a potential does not impact the thermally-averaged kinetic energy, which is \(k_B T/2\) in any case; quantum mechanically, the presence of a potential does alter the thermally-averaged kinetic energy of the system as well. 

\section{Rotating-wave approximation}
We now (briefly) comment on the rotating-wave approximation or the RWA, in the context of the dissipative quantum oscillator. For simplicity, we consider the linear-coupling model, which disregards the term proportional to \(x^2\) in \(H_{\rm SB}\) [Eq. (\ref{HSB})]. Introducing now, the lowering operators for the system and heat-bath oscillators as
\begin{equation}\label{abvariables}
a = \frac{m \omega_0 x + ip}{\sqrt{2 m \hbar \omega_0}}, \hspace{5mm} b_j = \frac{m_j \omega_j q_j + ip_j}{\sqrt{2 m_j \hbar \omega_j}},
\end{equation} with the raising operators being their hermitian conjugates, the interaction Hamiltonian, i.e., \(H_{\rm SB} = - x \sum_{j=1}^N c_j q_j\), reads
\begin{equation}\label{HBFC}
H_{\rm SB} = \sum_{j=1}^N \lambda_j (a b_j + a b_j^\dagger + a^\dagger b_j + a^\dagger b_j^\dagger), \hspace{3mm} \lambda_j = \frac{\hbar c_j}{2 \sqrt{ m m_j \omega_0 \omega_j}}.
\end{equation} We now invoke the RWA, which involves eliminating the rapidly-oscillating terms, which in this case are the ones proportional to \(a^\dagger b_j^\dagger\) and \(a b_j\) \cite{GSA,RWA,RWA1}. Thus, after implementing the RWA, the full Hamiltonian now reads 
\begin{eqnarray}\label{HL|}
H^{\rm RWA} &=& \hbar \omega_0  \bigg(a^\dagger a + \frac{1}{2} \bigg) + \sum_{j=1}^N \hbar \omega_j \bigg( b_j^\dagger b_j + \frac{1}{2}\bigg) \\
&& +  \sum_{j=1}^N \lambda_j (a^\dagger b_j + a b_j^\dagger). \nonumber
\end{eqnarray} 
In \cite{GSA}, a corresponding Fokker-Planck equation was obtained for the dissipative oscillator in the RWA, starting with a master equation. The resulting Langevin equations read as
\begin{eqnarray}
\dot{x}(t) &=& - \gamma x(t) + \frac{p(t)}{m} + f_x (t), \label{rwaana} \\
 \dot{p}(t) &=& - \gamma p(t) - m \omega_0^2 x(t) + f_p(t), \label{rwother}
\end{eqnarray} where \(f_{x,p}(t)\) are Gaussian noises with zero mean and the following two-point correlations: 
\begin{eqnarray}
\langle \{ f_x(t), f_x(t') \} \rangle = \frac{2\gamma\hbar}{ m \omega_0} \coth\bigg(\frac{\hbar\omega_0}{2k_BT}\bigg) \delta(t-t') , \\
\langle \{ f_p(t), f_p(t') \} \rangle = 2m\gamma\hbar \omega_0 \coth\bigg(\frac{\hbar\omega_0}{2k_BT}\bigg) \delta(t-t') .
\end{eqnarray}
Notice that Eq. (\ref{rwaana}) seems to present an anomaly, for we expect \(\dot{x} = p/m\). We now explain the origin of this supposed anomaly. Transforming back to the oscillator variables \((x,p)\) and \((q_j,p_j)\) by inverting Eq. (\ref{abvariables}), the RWA Hamiltonian [Eq. (\ref{HL|})] reads \cite{QLE}
\begin{eqnarray}
H^{\rm RWA} &=& \frac{p^2}{2m} + \frac{m\omega_0^2 x^2}{2} + \sum_{j=1}^{N} \Bigg[\frac{p_j^2}{2m_j}+\frac{1}{2}m_j\omega_j^2q_j^2\Bigg] \nonumber \\
&& + \frac{x}{2} \sum_{j=1}^N c_j q_j + \frac{p}{2 m \omega_0} \sum_{j=1}^N \frac{c_j p_j}{m_j \omega_j} .
\end{eqnarray}
Comparing this with the original linear-coupling Hamiltonian (the one without RWA being invoked), one finds that the RWA essentially introduces a momentum-momentum coupling between the system and the bath (see \cite{momentum1,momentum2,momentum} for details on momentum-momentum coupling), as if the coordinate-coordinate coupling has got split into half; one half still holds on to coordinate-coordinate coupling, while the other half now assumes the form of momentum-momentum coupling. In other words, the RWA `mixes' pure coordinate-coordinate coupling with momentum-momentum coupling. The resulting Heisenberg equations for the system operators read
\begin{eqnarray}
\dot{x}(t) &=& \frac{p(t)}{m} + \frac{1}{2 m \omega_0} \sum_{j=1}^N \frac{c_j p_j(t)}{m_j \omega_j}, \label{dotxrwa} \\
\dot{p}(t) &=& - m \omega_0^2 x(t) - \frac{1}{2} \sum_{j=1}^N c_j q_j(t). 
\end{eqnarray}
The RWA therefore, leads to a `new dissipation channel' (within \(\dot{x}\)) and this explains the appearance of explicit drag and noise forces in Eq. (\ref{rwaana}). Taking \(\gamma << \omega_0\), it follows that Eqs. (\ref{rwaana}) and (\ref{rwother}) consistently describe the dissipative quantum oscillator \cite{GSA}. It may be noted that Eq. (\ref{rwaana}) indicates that the Ehrenfest's theorem does not hold. We remark in passing that for Langevin equations obtained in the weak-coupling limit, it may only be justified to consider an underdamped regime; this is in contrast to strong-coupling situations (see also \cite{virial2}). \\

It should further be remarked that Eqs. (\ref{rwaana}) and (\ref{rwother}) describe a weak-coupling situation, being obtained under a Born-Markov approximation \cite{GSA}. However, the Hamiltonian given in Eq. (\ref{HL|}) does not yet correspond to a weak-coupling limit. The condition for relatively `weak' coupling is often additionally included within the RWA in the context of quantum optics, leading to the emergence of the Jaynes-Cummings Hamiltonian [Eq. (\ref{HL|})] from the Rabi Hamiltonian, the latter having an interaction Hamiltonian that goes as Eq. (\ref{HBFC}). It then ensures that both the Hamiltonians describe the same physics in a perturbative expansion involving terms of the higher order; terms such as \(a^\dagger b_j^\dagger\) and \(a b_j\) do not correspond to real (number conserving) processes and do not contribute in the first order of the perturbation theory.

\section{Discussion}
In this short note, we have described the implementation of the Born and Markov approximations in the context of the QLE, focusing on the simple case where the system is a harmonic oscillator. It was argued that in the general case, the formulation of the QLE requires the system and the bath to be coupled at the initial instant, i.e., at \(t = 0\), consistent with a strong-coupling scenario. Although we have worked with the (one-dimensional) independent-oscillator model, the fact that there are non-trivial system-bath correlations at the initial time can also be observed in other models such as the gauge-invariant model of a charged oscillator coupled to a heat bath of independent oscillators via momentum variables \cite{momentum}. \\

The Born and the Markov approximations are then implemented as follows. For the former, we take the system-bath coupling to be vanishing, by considering the limit \(\gamma \rightarrow 0^+\) in the correlation functions computed from the QLE. Here, we must mention a word of caution; while computing certain correlation functions such as that leading to the magnetic moment in the case of dissipative diamagnetism \cite{sdg1,kaur3}, it is important to perform the integral first, and then invoke the limit of taking the system-bath coupling to zero. This ensures that the boundary effects, encoded within the confining potential controlled by the parameter \(\omega_0\) are properly taken into account. However, as far as the mean kinetic and potential energies are concerned (at equilibrium), the limit $\gamma \rightarrow 0^+$ commutes with the integral over $\omega$. \\

For the Markov approximation, one assumes that the bath correlations fall off really fast, over time. This essentially leads to an approximately Markovian noise [Eq. (\ref{fMarkov})], with correlation \( \langle \lbrace f(t), f(t') \rbrace \rangle \approx  \Gamma \delta(t-t')\), where \(\Gamma = 2m\gamma\hbar \omega_0 \coth\big(\frac{\hbar\omega_0}{2k_BT}\big)\). Notice that \(\hbar\) appears explicitly within \(\Gamma\), suggesting that the problem is still quantum mechanical; in this special limit, the noise reduces to a Markovian one, unlike the general case in which even (strict) Ohmic dissipation does not ensure a Markovian noise \cite{QLE}. It is imperative to mention a related development \cite{HvsM}, wherein the authors discuss the Born and Markov approximations in the context of both Langevin and master equations. However, our approach differs from that of \cite{HvsM} because in the latter, the Langevin equations are considered in a weak-coupling scenario by taking a factorized initial density matrix; on the other hand, in this paper, we have started with the quantum Langevin at strong system-bath coupling, with the main result being the transit between the regimes of strong and weak coupling. We conclude by remarking that although the factor of \(\gamma\) drops from stationary (equilibrium) and equal-time correlation functions in the weak-coupling limit [Eqs. (\ref{Cxweak1}), (\ref{Cxdotweak1}), (\ref{xxeq1}), (\ref{xxeq})], it plays an important role in out-of-equilibrium results, even for weak coupling (see for example \cite{GSAW}). \\

\textbf{Acknowledements:} A.G. thanks Malay Bandyopadhyay, Jasleen Kaur, Shamik Gupta, and Akash Sinha for related discussions, and the Ministry of Education (MoE), Government of India, for support in the form of a Prime Minister's Research Fellowship (ID: 1200454). S.D. thanks the Indian National Science Academy for support through their Honorary Scientist Scheme. This work was initiated when S.D. was visiting IIT Bhubaneswar; he thanks Saroj Nayak, Malay Bandyopadhyay, and Samvit Mahapatra for hospitality.

\end{document}